\documentclass[10pt,conference]{IEEEtran}
\usepackage{amsmath,amssymb,amsfonts,latexsym,cite}
\usepackage[final]{graphicx}

\normalsize

\def\figref#1{Fig.~\ref{#1}}
\def\secref#1{Section~\ref{#1}}

\def\lemref#1{Lemma~\ref{#1}}
\def\thmref#1{Theorem~\ref{#1}}

\def\bdA{\mathbf{A}}
\def\bdB{\mathbf{B}}
\def\bdC{\mathbf{C}}
\def\bdH{\mathbf{H}}
\def\bdP{\mathbf{P}}
\def\bdQ{\mathbf{Q}}
\def\bdR{\mathbf{R}}
\def\bdT{\mathbf{T}}
\def\bdV{\mathbf{V}}
\def\bdd{\mathbf{d}}
\def\bdh{\mathbf{h}}
\def\bdy{\mathbf{y}}
\def\bdz{\mathbf{z}}
\def\mD{\mathcal{D}}
\def\mJ{\mathcal{J}}
\def\bdones{\mathbf{1}}

\def\bdLambda{\mathbf{\Lambda}}

\def\CN#1{\mathcal{CN}(#1)} % complex gaussian distribution
\let\td\tilde

\DeclareMathOperator{\diag}{diag}

\def\bydef{:=}
\def\a{{(a)}}
\def\b{{(b)}}
\def\c{{(c)}}
\def\set#1{\Bar{\mathcal{J}_#1}}
\def\ic{{i^c}}

\newtheorem{theorem}{Theorem}
\newtheorem{lemma}{Lemma}

\makeatletter\def\@oddfoot{\hfill\thepage\hfill}\makeatother

\begin{document}

\title{Interference Alignment and Degrees of Freedom Region
of Cellular Sigma Channel}

\author{ \normalsize
 \IEEEauthorblockN{Huarui Yin$^1$, Lei Ke$^2$, Zhengdao Wang$^2$}
 \small
 \IEEEauthorblockA{$^1$WINLAB, Dept of EEIS,
 Univ.~of Sci.~and Tech.~of China, Hefei, Anhui, 230027, P.R. China.
   email: yhr@ustc.edu.cn\\
 $^2$Dept.~of ECE, Iowa State University, Ames, IA 50011, USA.
   email: \{kelei,zhengdao\}@iastate.edu}
 }

\maketitle

\begin{abstract}
We investigate the Degrees of Freedom (DoF) Region of a cellular network,
where the cells can have overlapping areas. Within an overlapping area, the
mobile users can access multiple base stations. We consider a case where there
are two base stations both equipped with multiple antennas. The mobile
stations are all equipped with single antenna and each mobile station can
belong to either a single cell or both cells. We completely characterize the
DoF region for the uplink channel assuming that global channel state
information is available at the transmitters. The achievability scheme is
based on interference alignment at the base stations.
\end{abstract}
%
% \begin{keywords}
% Interference alignment, degrees of freedom region, cellular system, multiple
% antennas
% \end{keywords}

\section{Introduction}

Traditional cellular systems orthogonalize the channels such that signals sent
from different transmitters are supposed to be orthogonal, at least in the
ideal case, in time, frequency, or code dimensions. Such orthogonalization
yield technologies such as TDMA, FDMA, or CDMA. However, orthogonalization is
not the most efficient way of utilizing the available signal dimensions. This
can be clearly seen even in a simple two-user scalar Gaussian multiple access
channel: TDMA/FDMA is optimal only in one case where the bandwidth allocation
is proportional to the power available to the users. For any other
cases/rates, orthogonalization is strictly suboptimal
\cite[Fig.~15.8]{coth06}.

Non-orthogonal transmissions necessarily create interference at the receivers.
How to ``design'' or control such interference is the key to higher network
efficiency. While it is important to consider interference, the fully coupled
interference channel model may be too pessimistic. The reason is that for
cellular networks, the users well inside a cell have high signal to
interference ratio (SNR), and as a result interference can be neglected compared to
the useful signal. Whereas for users on the boundary of two or multiple cells,
the interference is comparable with the signal.
% It is also possible that
% signals from multiple cells are used all to deliver useful messages.

For interference networks, instead of trying to characterize the capacity
region completely, which is a difficult problem, the notion of degrees of
freedom (DoF) has been used to characterize how capacity scales with transmit
power as the SNR goes to infinity \cite{jafa07,mmk08,jash08}. The degrees of freedom
is also known as the multiplexing gain \cite{zhts03}. The interference alignment
for cellular network has been considered in \cite{suts08c} based on decomposable
channel, where it is shown that the interference free DoF can be achieved when the number of mobile stations increases. Practical usage of interference alignment in cellular network has been considered in \cite{trgu09c, suht10c}.

In this paper, we consider a cellular network that has overlapping cells.
Within an overlapping area, the mobile stations (MS) can access multiple base
stations (BS). This is a typical scenario in cellular communications. We consider a
simple case where there are only two base stations both equipped with multiple
antennas. We assume that the mobile stations are all equipped with single
antenna.

Our main contribution of the paper is the complete characterization of the
uplink DoF region for a cellular system with two base stations serving two
overlapping areas. The achievability scheme uses interference alignment at the
two base stations. As a special case of our result, we obtain the DoF region
of an X-network \cite{caja09} with single antenna at the transmitters
and multiple antennas at two receivers.

The rest of the paper is organized as follows. In \secref{sec.model}, we
present the system model of the problem considered. The statement of our main
result is presented in \secref{sec.main}. The proof of the converse is
presented in \secref{sec.converse}, and the achievability is established in
\secref{sec.achieve}. Finally, \secref{sec.conc} concludes the paper.

\section{System Model} \label{sec.model}

\begin{figure*}
\begin{minipage}{.49\textwidth}
\centering
\includegraphics[width=.6\linewidth]{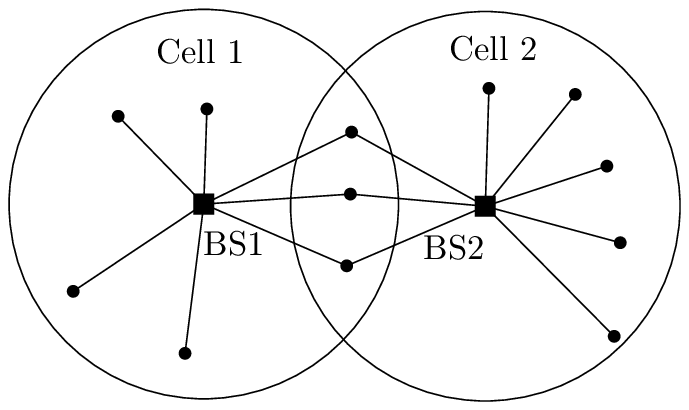}
\caption{Cellular network with overlapping cells.}
\label{fig.cell}
\end{minipage}
\begin{minipage}{.49\textwidth}
\centering
\includegraphics[width=.8\linewidth]{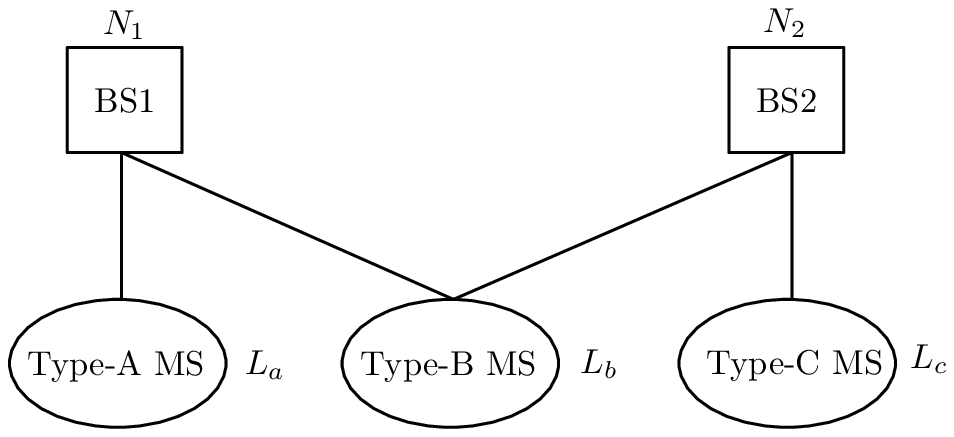}
\caption{The $\Sigma$ channel model.}
\label{fig.sigma}
\end{minipage}
\end{figure*}

We consider a typical communication scenario in wireless cellular system,
where there is an overlapping area between two cells; see \figref{fig.cell}.
For simplicity, we only consider a system of two base stations. We assume the
number of antennas at  two base stations are $N_1$ and $N_2$, respectively,
and all the mobile stations have single antenna. Depending on the locations of
the mobile stations, we divide the mobile stations into three groups:
\begin{enumerate}
\item Group A: mobile stations that communicate with BS $1$ only;
\item Group B: mobile stations that communicate with both BS $1$ and BS $2$;
\item Group C: mobile stations that communicate with BS $2$ only.
\end{enumerate}

Denote the number of mobile stations in different groups as $L_a$, $L_b$ and
$L_c$, respectively. We consider the uplink transmission, where mobile
stations in Group A or Group C each have only one independent message and
mobile stations in Group B generate independent messages for both base
stations. Therefore, the total number of messages is $L=L_a+2L_b+L_c$. We term such a channel as $\Sigma$ channel, due to the
resemblance, see \figref{fig.sigma}.
Let $t$ be time index. The
channel from the mobile station $j$ in Group A to BS~1 is denoted as
$\bdh_{1j}^\a(t) \in \mathbb{C}^{N_1\times 1}, 1\leq j\leq L_a$. The channel from
the mobile station $j$ in Group B to BS $i$ is denoted as $\bdh_{ij}^\b
(t)\in\mathbb{C}^{N_i\times 1}, i\in\{1,2\}, 1\leq j\leq L_b$. The channel from
the mobile station $j$ in Group C to BS~2 is denoted as $\bdh_{2j}^\c
(t)\in\mathbb{C}^{N_2\times 1}, 1\leq j\leq L_c$. The channel coefficients
in different time instants are all independently and identically generated from
some continuous distribution whose minimum and maximum values are finite.
When $i\in \{1,2\}$ denote the
index for BS, we use $\ic\bydef 3-i$ to refer to the index of the other BS.

Denote the messages and the corresponding transmitted signals from the three
types of mobile stations as
\begin{align*}
&W_{1j}^\a, x^\a_{1j}(t), 1\leq j\leq L_a \\
&W^\b_{ij},i \in \{1,2\}, 1\leq j\leq L_b, x^\b_{j}(t),1\leq j \leq L_b \\
&W^\c_{2j}, x^\c_{2j}(t), 1\leq j\leq L_c,
\end{align*}
respectively. The received signals of BS 1 and 2 are as follows
\begin{align}
\bdy_1&(t)=\sum_{j=1}^{L_a}\bdh_{1j}^\a (t)x_{1j}^\a(t)+
  \sum_{j=1}^{L_b}\bdh_{1j}^\b (t)x^\b_{j}(t)+\bdz_1(t),\\
\bdy_2&(t)=\sum_{j=1}^{L_c}\bdh_{2j}^\c
(t)x_{2j}^\c(t)+\sum_{j=1}^{L_b}\bdh_{2j}^\b (t)x^\b_{j}(t)+\bdz_2(t),
\end{align}
where $\bdz_i(t)\in \mathbb{C}^{N_i\times 1}, i\in \{1,2\}$ are the additive
$\CN{0,1}$ noise.

The power of all transmitted signals is limited to $P$. The achievable rates of
the messages are denoted as $R_{1j}^\a(P), 1\leq j\leq L_a$, $R^\b_{ij}(P), i
\in \{1,2\}, 1\leq j\leq L_b$, and $R^\c_{2j}(P), 1\leq j\leq L_c$,
respectively. The associated DoF of the messages are
\begin{align*}
d_{1j}^\a&=\lim_{P\rightarrow \infty}\frac{R_{1j}^\a(P)}{\log(P)},
  1\leq j\leq L_a, \\
d^\b_{ij}&=\lim_{P\rightarrow \infty}\frac{R^\b_{ij}(P)}{\log(P)},
  i \in \{1,2\}, 1\leq j\leq L_b, \\
d^\c_{2j}&=\lim_{P\rightarrow \infty}\frac{R^\c_{2j}(P)}{\log(P)},
  1\leq j\leq L_c.
\end{align*}
 Let $\bdR_L(P)\in\mathbb
R^+_L$ denote the vector containing all the rates at power $P$, and $\bdd_L\in
\mathbb R^+_L$ denote the vector containing all the DoF. Let $C(P)\subset
\mathbb R^+_L$ denote the capacity region of the system, which contains all
the rate tuples $\bdR_L(P)$ such that the probability of error at all
receivers can approach zero as the coding length tends to infinity. The DoF
region of the $\Sigma$ channel is the collection of all the DoF points
\begin{multline*}
\mD \bydef \left\{
  \bdd_L\in \mathbb R^+_L:
  \exists\bdR_L(P)\in C(P) \text{ such that } \right. \\
  \left. \bdd_L=\lim_{P\to\infty}\frac{\bdR_L(P)}{\log(P)}
\right \}.
\end{multline*}

\section{Main Result} \label{sec.main}

The main result of our paper is the following theorem.
\begin{theorem} \label{thm.sgmrg}
The DoF region of $\Sigma$ channel with global channel state information
at transmitters is specified by the following inequalities
\begin{align}
  d^\a_{1j}\leq 1,\quad 1\leq j \leq L_a \label{eq.sgmrg.a};\\
  d^\b_{1j}+d^\b_{2j} \leq 1, \quad 1\leq j \leq L_b \label{eq.sgmrg.b};\\
  d^\c_{2j}\leq 1, \quad 1\leq j \leq L_c \label{eq.sgmrg.c};\\
  \sum_{j=1}^{L_a}d^\a_{1j}+\sum_{j=1}^{L_b}d^\b_{1j}+\sum_{j\in
    \mathcal{J}_2}d^\b_{2j} \leq N_1, \notag \\
    \quad \forall \mathcal{J}_2\subseteq \left\{1,2,\cdots,L_b\right\},
      |\mathcal{J}_2|\leq N_1; \label{eq.sgmrg.mac1}\\
  \sum_{j=1}^{L_c}d^\c_{2j}+\sum_{j=1}^{L_b}d^\b_{2j}+
    \sum_{j\in\mathcal{J}_1}d^\b_{1j}\leq N_2,\notag \\
    \quad \forall \mathcal{J}_1\subseteq \left\{1,2,\cdots,L_b\right\},
    |\mathcal{J}_1|\leq N_2. \label{eq.sgmrg.mac2}
\end{align}
\end{theorem}

\section{Proof of The Converse} \label{sec.converse}

We prove the converse part of \thmref{thm.sgmrg} in this section. The first
three inequalities \eqref{eq.sgmrg.a}--\eqref{eq.sgmrg.c} are due to the fact
that all the mobile stations have single antennas. To show the upper bound
\eqref{eq.sgmrg.mac1}, we first divide the mobile stations in Group B into two
parts. We treat the mobile stations of Group B whose indices in $\mJ_2$ as a
super user $S_b$, whose messages are $W_{1j}^\b,   W_{2j}^\b, j\in\mJ_2$ and
number of antennas is $|\mJ_2|$. We treat the remaining mobile stations in Group B
and the mobile stations in Group A as another super user $S_a$. In addition,
we assume all the messages $W_{2j}^\b=0, j\notin \mJ_2$. Therefore, the super
user $S_a$ has $L_a+L_b-|\mJ_2|$ antennas and its messages are $W_{1j}^\a, 1\leq
j\leq L_a$ and $W_{1j}^\b, 1\leq j\leq L_b,j\notin \mJ_2$. We further assume
that the mobile stations within $S_a$ and $S_b$ can fully cooperate with each
other if they belong to the same super user set. Then $S_a$, $S_{b}$, BS 1 and
BS 2 form a  multiple-input and multiple-output Z interference channel. Since cooperation does no harm to the
degrees of freedom, the inequality \eqref{eq.sgmrg.mac1} holds based on
\cite[Corollary~1]{jafa07}. The inequality \eqref{eq.sgmrg.mac2} can be proved
similarly.

\section{Achievability of The DoF Region} \label{sec.achieve}

We here give the achievability scheme of the DoF region of $\Sigma$ channel
based on interference alignment over the time expansion channel. The proof is
built upon the alignment scheme proposed in \cite{goja08a}.

\subsection{Time expansion modelling}
For any rational DoF point $\bdd_L$ within $\mD$ and satisfying
\eqref{eq.sgmrg.a}--\eqref{eq.sgmrg.mac2}, we can choose a positive integer
$\mu_0$, such that
\begin{align}
\mu_0\bdd_L&\in\mathbb{Z}_+^L,
\end{align}
where $\mathbb{Z}_+^L$ denotes the set of $L$-dimensional non-negative
integers. For any irrational DoF point in the DoF region, we can always
approximate it as a rational point with arbitrarily small error. Denote
$\mu_n$ as the duration in number of symbols of the time expansion. Here and
after, we use the $\tilde\ $ notation to denote the time expanded signals.
Hence, $\tilde
\bdH_{ij}^\b=\diag(\bdh_{ij}^\b(1),\bdh_{ij}^\b(2),\dots,\bdh_{ij}^\b(\mu_n))$, which
is a size $N_i \mu_n\times \mu_n$ block diagonal matrix. Matrices $\td
\bdH_{1j}^\a$ and $\td\bdH_{2j}^\c$ can be similarly defined.

For BS 1, we have the following two cases:
\begin{enumerate}
\item When $L_b> N_1$, we will align $L_b-N_1$ interference messages at BS 1.
For any DoF point $\bdd_L$ within $\mD$, let $\set2$ denote a set containing
the indices of mobile stations in Group B such that $|\set2|=N_1$ and
\begin{align*}
\!\sum_{j\in \set2}d_{2j}^\b \geq\! \sum_{j\in \mJ_2}d_{2j}^\b,\! \quad
  \forall \mJ_2\!\subseteq\! \left\{1,2,\cdots,L_b\right\},
    |\mathcal J_2|\leq N_1.
\end{align*}
Furthermore, let
\begin{align}
\delta_2=\min\left\{j \bigg| j\in\set2, \text{and }
  d_{2j}^\b=\min_{k\in\set2}{d_{2k}^\b}  \right \}.
\end{align}
As we will see, the interference messages $W_{2j}^\b, j\in \set2$ will span
the interference space at BS 1 after going through the time-varying channel.
Among all these messages $W_{2\delta_2}^\b$ is the message having smallest
DoF. For any other messages $W_{2j}^\b, j\notin \set2$, its DoF must be less
than $d_{2\delta_2}^\b$. We will align these messages to message
$W_{2j}^\b, j\in \set 2$ at BS1.

\item When $L_b\leq N_1$, choose $\mathcal{J}_2=\{1,\dots,L_b\}$,
\eqref{eq.sgmrg.mac1} becomes
\begin{align}
     \sum_{j=1}^{L_a}d^\a_{1j}+\sum_{j=1}^{L_b}(d^\b_{1j}+d^\b_{2j})&\leq
     N_1
\end{align}
which suggests that all the messages are decodable at BS~1. Therefore there is
no need to do interference alignment for BS 1.
\end{enumerate}

Similarly. For BS 2, we have
\begin{enumerate}
\item when $L_b> N_2$, let $\set1$ denote the set containing the indices of
mobile stations in Group B such that $|\set1|=N_2$ and
\begin{equation}
\!\sum_{j\in \set1}d_{1j}^\b \geq\! \sum_{j\in \mJ_1}d_{1j}^\b,\!  \quad
\forall \mathcal{J}_1\!\subseteq\! \left\{1,2,\cdots,L_b\right\},
|\mathcal{J}_1|\leq N_2. \notag
\end{equation}
In addition, let
\begin{align}
\delta_1=\min\left\{j \bigg| j\in\set1, \text{and }
d_{1j}^\b=\min_{k\in\set1}{d_{1k}^\b}  \right \}.
\end{align}
\item when $L_b\leq N_2$, there is no need to do alignment.
\end{enumerate}

Let $\Gamma_1=N_1\max(L_b-N_1,0)$ and $\Gamma_2=N_2\max(L_b-N_2,0)$. We shall
see that they are the numbers of alignment constraints that mobile stations in
Group B need to satisfy in order to align interference at BS 1 and BS 2,
respectively.

We propose to use $\mu_n=\mu_0(n+1)^{\Gamma_1+\Gamma_2}$ fold time expansion,
where $n$ is a positive integer. Specifically, we want to achieve the
following DoF over $\mu_n$ slots
\begin{align}
\bar d_{1j}^\a&=\mu_0n^{\Gamma_1+\Gamma_2}d_{1j}^\a, 1\leq j\leq L_a,
  \label{eq.dof.1a}\\
\bar d_{1j}^\b&=\mu_0n^{\Gamma_1+\Gamma_2}d_{1j}^\b, 1\leq j\leq L_b, j\notin
\set1, \label{eq.dof.1bc}\\
\bar d_{1j}^\b&=\mu_0n^{\Gamma_1}(n+1)^{\Gamma_2}d_{1j}^\b, 1\leq j\leq L_b,
  j\in \set1, \label{eq.dof.1b}\\
\bar d_{2j}^\b&=\mu_0n^{\Gamma_1+\Gamma_2}d_{2j}^\b, 1\leq j\leq L_b, j\notin
\set2, \label{eq.dof.2bc}\\
\bar d_{2j}^\b&=\mu_0(n+1)^{\Gamma_1}n^{\Gamma_2}d_{2j}^\b, 1\leq j\leq L_b,
  j\in \set2, \label{eq.dof.2b}\\
\bar d_{2j}^\c&=\mu_0n^{\Gamma_1+\Gamma_2}d_{2j}^\c,
  1\leq j\leq L_c.\label{eq.dof.2c}
\end{align}
Therefore, when $n\rightarrow \infty$, the desired DoF point $\bdd_L$ can be
achieved. The key is to design the beamforming column sets for mobiles
stations in Group B such that the DoF \eqref{eq.dof.1a}--\eqref{eq.dof.2c} can
be achieved over $\mu_n$ slots.

\subsection{Beamforming and interference alignment}

When $L_b\leq N_1$, no interference alignment is needed at BS 1, in which case
we choose $\td\bdV_{2j}^\b$ to be a size $\mu_n\times \bar d_{2j}^\b$ random
full rank matrix. Similarly, when $L_b\le N_2$, we choose $\td\bdV_{1j}^\b$ to
be a size $\mu_n\times \bar d_{1j}^\b$ random full rank matrix. In the
following, we assume that $L_b> \max(N_1,N_2)$ and will design the beamforming
matrices.

We choose the beamforming column sets of mobile stations in Group B whose
indices belong to $\set1$ and $\set2$ to have the following forms
\begin{align}
\td\bdV_{1j}^\b&=[\td\bdP_{11},\td\bdQ_{1j}^\b], \quad j \in \set1,
  \label{eq.bf.groupb1}\\
\td\bdV_{2j}^\b&=[\td\bdP_{21},\td\bdQ_{2j}^\b], \quad j \in \set2,
  \label{eq.bf.groupb2}
\end{align}
where $\td\bdQ_{1j}^\b$ is a size $\mu_n\times (\bar d_{1j}^\b-\bar
d_{1\delta_1}^\b)$ random matrix, and $\td\bdQ_{2j}^\b$ is a size $\mu_n\times
( \bar d_{2j}^\b-\bar d_{2\delta_2}^\b)$ random matrices. Obviously
\begin{align}
\td\bdV_{i\delta_i}^\b =\td\bdP_{i1},\quad i\in\{1,2\}.
\end{align}
The two matrices $\td\bdP_{11}$ and $\td\bdP_{21}$ are structured and will be
determined later. In our design \eqref{eq.bf.groupb1} and
\eqref{eq.bf.groupb2}, all the messages $W_{1j}^\b, j\in\set1$ sharing part of
the same beamforming columns which is $\td\bdP_{11}$. Similarly, all the
messages $W_{2j}^\b, j\in\set2$ sharing part of the same beamforming columns
which is $\td\bdP_{21}$.

For $i=1,2$, by the definition of $\set i$, the DoF of message $W_{ij}^\b,
j\notin \set i$ is at most the same as that of $W_{i \delta_i}^\b$. It would
therefore be sufficient if we could design a beamforming column set, denoted
as $\td\bdP_{i2}$, to be used by mobile station $j\notin \set i$, which is
able to deliver message with DoF $\bar d_{i\delta_i}^\b$.

Denote the set of elements of $\bar\mJ_{i}$ as
$\{\beta_{i,1},\beta_{i,2},\dots,\beta_{i,N_{i^c}}\}$. As the channel are random
generated, the following channels
\begin{align}
\td\bdH^{(1)}=[\td\bdH^\b_{1\beta_{2,1}},\td\bdH^\b_{1\beta_{2,2}},
  \dots,\td\bdH^\b_{1\beta_{2,N_1}}],\\
  \td\bdH^{(2)}=[\td\bdH^\b_{2\beta_{1,1}},\td\bdH^\b_{2\beta_{1,2}},
  \dots,\td\bdH^\b_{2\beta_{1,N_2}}],
\end{align}
have full rank with probability 1. As observed in \cite{goja08a}, it is
impossible to align interference message $W_{ij}^\b, j\notin \set i$ from
mobile station $j$ to only one interference message $W_{ik}^\b, k\in \set{i}$ at
BS $\ic$, because the channel between any $N_\ic$ mobile stations to BS $\ic$
are linear independent with probability one. However, we can choose
\begin{align}
\begin{split}
\!\!\!\td\bdH_{1j}\td\bdP_{22}\prec
[\td\bdH^\b_{1\beta_{2,1}}\td\bdP_{21},\td\bdH^\b_{1\beta_{2,2}}\td\bdP_{21},
  \dots,\td\bdH^\b_{1\beta_{2,N_1}}\td\bdP_{21}],
\\
j\notin \set2,
\end{split} \label{eq.align.bs1}\\
\begin{split}
\!\!\!\td\bdH_{2j}\td\bdP_{12}\prec [\td\bdH^\b_{2\beta_{1,1}}\td\bdP_{11},
  \td\bdH^\b_{2\beta_{1,2}}\td\bdP_{11},\dots,
  \td\bdH^\b_{2\beta_{1,N_2}}\td\bdP_{11}], \\
j\notin \set1,
\end{split}
\label{eq.align.bs2}
\end{align}
so that the interference space at BS $\ic$ is not larger than that spanned by
messages in $\set i$.

We define the $\mu_n\times \mu_n$ matrices $\bdT_{l}^{(ij)}$, $l=1, 2, \ldots,
N_i$ according to the following
\begin{align}
\begin{bmatrix} \bdT_{1}^{(ij)} \\
\bdT_{2}^{(ij)} \\
\vdots\\
\bdT_{N_i}^{(ij)}
\end{bmatrix}= \left(\td\bdH^{(i)} \right)^{-1}\td\bdH_{ij},\quad
  j\notin \set {\ic}.
\end{align}
It has been shown in \cite{goja08a} that these $\bdT_{l}^{(ij)}$ matrices are
diagonal matrices. We also define the following block diagonal matrices
\begin{align}
\td\bdP^{(1)}=\diag(\underbrace{\td\bdP_{11},\td\bdP_{11},\dots,
  \td\bdP_{11}}_{\text{$N_2$ matrices} }),\\
\td\bdP^{(2)}=\diag(\underbrace{\td\bdP_{21},\td\bdP_{21},\dots,
  \td\bdP_{21}}_{\text{$N_1$ matrices} }).
\end{align}
It follows from \eqref{eq.align.bs1} and \eqref{eq.align.bs2} that
\begin{align}
\begin{bmatrix} \bdT_{1}^{(ij)} \\
\bdT_{2}^{(ij)} \\
\vdots\\
\bdT_{N_i}^{(ij)}
\end{bmatrix}\td\bdP_{\ic2}\prec  \td\bdP^{(\ic)}, \quad j\notin \set{{i^c}}.
\end{align}
Therefore, the alignment constraints for BS 1 and 2 are
\begin{align}
\bdT_{l}^{(1j)}\td\bdP_{22}\prec\td\bdP_{21}, \quad
  j\notin\set {2}, 1\leq l\leq N_1, \label{eq.align1} \\
  \bdT_{l}^{(2 j)}\td\bdP_{12}\prec\td\bdP_{11}, \quad
  j\notin\set {1}, 1\leq l\leq N_2,\label{eq.align2}
\end{align}
and the total number of constraints is $\Gamma_1$ and $\Gamma_2$, respectively.

Denote $B_1=\mu_0 n^{\Gamma_1}d_{1\delta_1}^\b$ and $B_2=\mu_0
n^{\Gamma_2}d_{2\delta_2}^\b$. The matrices
$\td\bdP_{11},\td\bdP_{12},\td\bdP_{21},\td\bdP_{22}$ are designed in
\eqref{eq.precoder.a}--\eqref{eq.precoder.d}.
\begin{figure*}
\centering
\begin{align}
\td\bdP_{11} &= \bigcup\limits_{m = 0}^{B_1 - 1}
  \left\{ \left( \prod\limits_{1\leq
  l \leq {N_2}, j \notin \set1} \left( {\bdT_l^{\left( {2j} \right)}} \right)^
  {\alpha _l^{(2j)}}  \right) \bdones_{\mu_n},
  \alpha _l^{\left( {2j} \right)} \in \left\{ {mn + m + 1,mn + m + 2, \cdots ,
    \left( {m + 1} \right)n + m + 1} \right\}\right\}  \label{eq.precoder.a} \\
\td\bdP_{12} &= \bigcup\limits_{m = 0}^{{B_1} - 1}
  \left\{ \left( \prod\limits_{1\leq l \leq {N_2},j \notin \set1}
    \left(\bdT_l^{(2j)} \right)^{\alpha _l^{(2j)}} \right) \bdones_{\mu_n},
    \alpha _l^{(2j)} \in \left\{ mn + m + 1,mn + m + 2, \cdots ,(m + 1)n + m
    \right\}\right\}  \\
\td\bdP_{21} &= \bigcup\limits_{m = 0}^{B_2 - 1}
  \left\{ \left( \prod\limits_{1\leq l \leq{N_1}, j \notin \set2}
    \left( \bdT_l^{(1j)} \right)^{\alpha _l^{(1j)}} \right)
    \bdones_{\mu_n},\alpha _l^{(1j)} \in \left\{
      mn + m + 1,mn + m + 2, \cdots ,( m + 1)n + m + 1 \right\}\right\}  \\
\td\bdP_{22} &= \bigcup\limits_{m = 0}^{B_2 - 1}
  \left\{ \left( \prod\limits_{1\leq l \leq N_1 ,j \notin \set2}
  \left( \bdT_l^{(1j)} \right)^{\alpha _l^{(1j)}}\right) \bdones_{\mu_n},
  \alpha _l^{(1j)} \in \left\{ mn + m + 1, mn + m + 2, \cdots ,
  (m + 1)n + m \right\}\right\} \label{eq.precoder.d}
\end{align}
\rule{.3\textwidth}{.5pt}
\end{figure*}

It can be verified that the number of columns of $\td\bdP_{11}$ and
$\td\bdP_{21}$ are $\mu_0 n^{\Gamma_1}(n+1)^{\Gamma_2}d_{1\delta_1}^\b$ and
$\mu_0(n+1)^{\Gamma_1} n^{\Gamma_2}d_{2\delta_2}^\b$, respectively. Therefore
messages $W_{1\delta_1}^\b$ and $W_{2\delta_2}^\b$ can achieve desired DoF
over $\mu_n$ slots when $n\rightarrow \infty$. It can be verified that if
message $W_{ij}^\b, j\notin \set i$ use $\td \bdP_{i2}$ as the beamforming
matrix, its signal will fall into the interference subspace spanned by
messages $W_{ij}^\b, j\in\set i$ at BS $\ic$. That is, all the alignment
conditions in \eqref{eq.align1} and \eqref{eq.align2} are satisfied.

Having specified the matrices $\td\bdP_{11}, \td\bdP_{12}, \td\bdP_{21},
\td\bdP_{22}$, we describe the beamforming matrices of all mobile stations in
the following.
\begin{enumerate}
\item For mobile station $j$ in Group B, if $j\in\set1$, it uses the
beamforming matrix in \eqref{eq.bf.groupb1} to transmit message $W_{1j}^\b$ to
BS 1. Mobile station $j\notin \set1$ randomly chooses $\bar d_{1j}^\b$ columns
of $\td \bdP_{12}$ as the beamforming matrix.

Similarly, if $j\in\set2$, mobile station $j$ uses the beamforming matrix in
\eqref{eq.bf.groupb2} to transmit message $W_{2j}^\b$ to BS 2. mobile station
$j\notin \set2$ randomly chooses $\bar d_{2j}^\b$ columns of $\td \bdP_{22}$
as the beamforming matrix.

\item Each mobile station $j$ in Group A randomly generates a $\mu_n\times
\bar d_{1j}^\a$ matrix as beamforming matrix $\td \bdV_{1j}^\a$.

\item Each mobile station $j$ in Group C randomly generates a $\mu_n\times
\bar d_{2j}^\c$ matrix as beamforming matrix $\td \bdV_{2j}^\c$.

\item All the entries of the random beamforming columns of mobile stations in
Group A, Group B and Group C are independently and identically generated from
some continuous distribution whose minimum and maximum values are finite.
\end{enumerate}

\subsection{Full Rankness} To guarantee that both base stations can decode the
desired messages, we first need to make sure that all the beamforming matrices
are full rank, which is easy to verify. In addition, we need to guarantee the
following conditions
\begin{enumerate}
\item $[\td \bdV_{1j}^\b, \td \bdV_{2j}^\b]$ has full column rank for any
mobile station $j$ in Group B.
\item The interference space and signal space are independent for both base
stations.
\end{enumerate}
The first condition is needed to guarantee that two messages of mobile
stations in Group B can be distinguished. It can be verified that $[\td
\bdP_{11},\td \bdP_{21}]$ has $\mu_0 n^{\Gamma_1} (n+1)^{\Gamma_2}d_{1\delta_1}^\b+\mu_0 (n+1)^{\Gamma_1} n^{\Gamma_2}
d_{2\delta_2}^\b$ columns, which may be larger than the
number $\mu_n$ of rows. However, the beamforming matrix for any mobile station
will contain a subset of the columns of $[\td \bdP_{11},\td \bdP_{21}]$, plus
possibly some additional random columns. Since $d_{1j}^\b+d_{2j}^\b\leq 1$,
$[\td \bdV_{1j}^\b, \td \bdV_{2j}^\b]$ is always a tall matrix (more rows than
columns). To establish its full rankness, note that each entry is a monomial,
the random variables that define the monomials are different for all rows due
to the time varying channel. In addition, in one row, the exponents of the
monomials are different. Therefore, the conditions in
\lemref{lem.random.fullrank} are satisfied and the sub-matrix has full rank.

We then need to validate the second condition, which is needed to guarantee
that the mobile stations can decode the messages that they are interested in.
We only show this for BS 1 as the same argument can be applied to BS 2 as
well. Define the following matrix
\begin{align}
\td\bdQ^{(2)}=\left [\begin{array}{c}
\diag(\td\bdQ_{2\beta_{2,1}}^\b ,
  \td\bdQ_{2\beta_{2,2}}^\b ,\dots,
  \td\bdQ_{2\beta_{2,N_1-1}}^\b) \\
\mathbf{0} \\
\end{array} \right],
\end{align}
\begin{figure*}
\abovedisplayskip -10pt \centering
\begin{align}
\bdLambda_1=\left[\underbrace{
  \td\bdH^\a_{11}\td\bdV_{11}^\a,\td\bdH^\a_{12}\td\bdV_{12}^\a,
  \dots,\td\bdH^\a_{1L_{a}}\td\bdV_{1L_{a}}^\a}_{\bdA},
\underbrace{
  \td\bdH^\b_{11}\td\bdV_{11}^\b,\td\bdH^\b_{12}\td\bdV_{12}^\b,\dots,
  \td\bdH^\b_{1L_{b}}\td\bdV_{1L_{b}}^\b}_{\bdB},
\underbrace{\td\bdH^{(1)}\td\bdQ^{(2)},\td\bdH^{(1)}\td\bdP^{(2)}}_{\bdC}\right]
\label{eq.space.abc}
\end{align}
\rule{.8\textwidth}{.5pt}
\end{figure*}%
where $\mathbf{0}$ is an all zero matrix.
When $L_b>N_1$ we would like to show that the matrix in \eqref{eq.space.abc}
has full column rank. Here $\bdA$ and $\bdB$ correspond to the signal part,
while $\bdC$ corresponds to the interference space generated by the mobile
stations whose indices are in $\set2$. The number of columns of $\bdA$ is
$\mu_0n^{\Gamma_1+\Gamma_2}\sum_{j=1}^{L_a}d_{1j}^\a$. The number of columns
of $\bdB$ is
$\mu_0n^{\Gamma_1+\Gamma_2}\sum_{j=1,j\notin\set1}^{L_b}d_{1j}^\b+
\mu_0n^{\Gamma_1}(n+1)^{\Gamma_2}\sum_{j\in\set1}d_{1j}^\b$. The number of
columns of $\bdC$ is
$\mu_0(n+1)^{\Gamma_1}n^{\Gamma_2}\sum_{j\in\set2}d_{2j}^\b$. Therefore, the
number of columns of $\bdLambda_1$ is less than
$\mu_0(n+1)^{\Gamma_1+\Gamma_2}N_1$ due to \eqref{eq.sgmrg.mac1}. Hence,
$\bdLambda_1$ is a tall matrix. We can see that the conditions of
\lemref{lem.random.fullrank} still hold due to the following reasons: 1) All
the elements of $\bdLambda_1$ are monomials of different random variables. 2)
The random variables of different rows are different. 3) The random variables
of $\bdV_{1j}^\a$ do not appear in $\bdB$ and $\bdC$. The random variables in
$\td\bdH^\b_{1j}$ do not appear in $\bdC$. The random variables in
$\td\bdP^{(2)}$ and $ \td\bdQ^{(2)}$ do not appear in $\bdA$ and $\bdB$.
Therefore, the associated exponents of monomials in one row differs at least
by one. Based on this, we conclude that $\bdLambda_1$ has full column rank and
the signal space is independent from interference space at BS 1. The proof of
the achievability is then complete.

\section{Conclusion} \label{sec.conc}

In the paper we proposed a Sigma ($\Sigma$) channel model for cellular
communication networks with overlapping cell areas. We allowed the base
stations to have multiple antennas and the mobile stations to have single
antennas. We derived the degrees of freedom region for the uplink
communication in the simple cellular network of two base stations, under the
assumption that global channel state information is available at the
transmitters. The achievability scheme is based on beamforming at the mobile
stations and interference alignment at the base stations.

\section*{Appendix}

\begin{lemma}\cite[Lemma~1]{caja09} \label{lem.random.fullrank}
Consider an $M\times M$ square matrix $\bdA$ such that $a_{ij}$, the elements
in the $i$th row and $j$th column of $\bdA$, is of the form
\begin{align}
a_{ij}=\prod_{k=1}^K\left(x_i^{[k]}\right)^{\alpha_{ij}^{[k]}}
\end{align}
where $x_i^{[k]}$ are random variables and all exponents are integers,
$\alpha_{ij}^{[k]}\in\mathbb{Z}$. Suppose that
\begin{enumerate}
\item $x_i^{[k]}|\{x_{i'}^{[k']}, \forall (i,k)\neq (i',k')\}$ has a
continuous cumulative probability distribution.
\item $\forall i,j,j'\in\{1,2,\dots,M\}$ with $j\neq j'$
\begin{align}
\left(\alpha_{ij}^{[1]},\alpha_{ij}^{[2]},\dots,\alpha_{ij}^{[K]}   \right)\neq
\left(\alpha_{ij'}^{[1]},\alpha_{ij'}^{[2]},\dots,\alpha_{ij'}^{[K]}   \right).
\end{align}
In other words, each random variable has a continuous cumulative probability
distribution conditioned on all the remaining variables. Also, any two terms
in the same row of the matrix $\bdA$ differ in at least one exponent.
\end{enumerate}
Then, the matrix $\bdA$ has a full rank of $M$ with probability 1.\hfill \QED
\end{lemma}

\linespread{1.1}\normalsize
\bibliographystyle{IEEEtran}
\bibliography{sigma}

\end{document}